\documentclass{elsart}
\usepackage[all]{xy}
\usepackage{graphicx}
\journal{Physics Letters B}

\newcommand{\be}{\begin{equation}}
\newcommand{\ee}{\end{equation}}
\newcommand{\bea}{\begin{eqnarray}}
\newcommand{\eea}{\end{eqnarray}}

\def\lesssim{\mathrel{\hbox{\rlap{\hbox{\lower4pt\hbox{$\sim$}}}\hbox{$<$}}}}
\def\gtrsim{\mathrel{\hbox{\rlap{\hbox{\lower4pt\hbox{$\sim$}}}\hbox{$>$}}}}

\begin{document}

\begin{frontmatter}
\title{Probing Light Pseudoscalar Particles Using Synchrotron Light}

\author{Alex G. Dias and G. Lugones}
\address{Universidade Federal do ABC,   Centro de Ci\^encias Naturais e Humanas \\
Rua Santa Ad\'elia, 166, 09210-170, Santo Andr\'e,  Brazil}

\begin{abstract}
The need for purely laboratory-based light pseudoscalar particles searches has been emphasized many times in the literature, since astrophysical bounds on these particles rely on several assumptions to calculate the flux produced in stellar plasmas.
In this paper we study the use of light from synchrotron accelerators as a source for
a photon regeneration experiment also know as ``light shining through a wall''.
Such an experiment can significantly improve present limits on the pseudoscalar particle mass and the pseudoscalar-photon coupling  constant obtained from laser experiments. This is possible even using a small number of powerful  magnets (B $\sim$ 10 T), due to the large incident photon flux. On the other hand, the use of a broadband incident photon-beam instead of infrared or optical lasers allows a significant improvement in the mass reach of the experiment (it is possible to test masses up to $0.01$ eV  without a drop in sensitivity).
Large, but still feasible, configurations can explore in a quite model-independent
way a large part of the parameter space examined by solar searches and HB stars in globular clusters. Additionally, the proposal may be useful for testing string motivated effective theories containing light and weakly interacting particles.

\end{abstract}


\end{frontmatter}

\date{}

\maketitle

%
\textit{Introduction:} For more than two decades several experiments
were designed for detecting phenomena associated with fundamental
pseudoscalar particles. These particles are predicted by many
theories beyond the Standard Model. For example, the existence of
particular trilinear terms like
\bea
{\cal{ L}}_{a-f} = g_{_P}\, a\, \bar{f} \gamma_5 f \, ,
\label{Laf}
\eea
composed by two charged fermion fields $f$ and a pseudoscalar $a$
in the interaction lagrangian, can generate a dimension five
effective operator ${\cal L}_{{a\gamma}}= -\frac{1}{4}g_{{a\gamma}}\,
a \, F_{\mu\nu}{\tilde{F}}^{\mu\nu}= g_{{a\gamma}}\, a\, {\bf E \,.\, B}$.
A remarkable example
is the solution of the Strong CP problem proposed by Peccei and
Quinn \cite{pq}. In this solution a pseudo-Goldstone boson known as
the axion \cite{axion} arises from the spontaneous breakdown of a
global symmetry. The axion may have tree level
interactions like Eq. (\ref{Laf}), so an electromagnetic coupling
of the form ${\cal L}_{{a\gamma}}$ could be obtained. The generic
photon-pseudoscalar interaction arising in such models is a
powerful tool for testing this new physics experimentally/observationally
(e.g. exploiting the fact that in the presence of a strong magnetic field,
a photon of frequency $\omega$ may oscillate into a pseudoscalar
particle of mass $m < \omega$, and viceversa).

The effective Lagrangian describing a photon-pseudoscalar system can
be wri\-tten as:
\bea
{\cal{ L}}_{\mathrm{eff}} & = &  -\frac{ 1}{4} F_{\mu\nu} F^{\mu\nu}
+\frac{ 1}{2 }\partial_\mu a \partial^\mu a - \frac{ m^2}{2 }a^2
-\frac{g_{{a\gamma}}}{4} \, a \, F_{\mu\nu} {\tilde{F}}^{\mu\nu},
\label{Leff}
\eea
which may be considered as a general Lagrangian containing the lowest
dimensional operators involving a pseudoscalar and the electromagnetic
field. Then, a\-ssu\-ming that there is no relation between
$g_{{a\gamma}}$ and $m$, the results derived from Eq. (\ref{Leff})
can be used to constrain these parameters independently.
These pseudoscalar particles are usually called \textit{axion-like}
particles.
Alternatively, a model dependent relationship between
$g_{{a\gamma}}$ and $m$ can be considered. For the so called
\textit{axion models} the coupling constant and the axion mass can
be related by an energy scale $f_a$ suppressing the
interaction and linked with the global symmetry breakdown. Thus,
\bea
g_{{a\gamma}} = \xi \frac{ \alpha }{2 \pi f_a},\,\,\,\,\,\,\,\,\,\,
m  \approx \frac{\Lambda_{QCD}^2}{f_a},
\label{coupmass}
\eea
where $\xi \sim 1$ is a model dependent factor, $\alpha$ is the
electromagnetic coupling, and $\Lambda_{QCD}$ is the usual QCD scale.
The predicted linear relationship between the coupling constant and
the axion mass allows to exclude models with basis on the axion-photon
conversion probability, which is a function of the parameters in
Eq. (\ref{coupmass}) \cite{sikivie83}.
%
%


The accumulated results from experiments along the years together
with several astrophysical constraints have tested a considerable
portion of the $g_{a\gamma}$ versus $m$  parameter-space
\cite{battesti-axionrev07,asztalos-axionrev06,Raffeltbook}. An important
observational limit on $g_{{a\gamma}}$ comes from the search of solar
axions where the bound $ g_{{a\gamma}}< 8.8 \times 10^{-11}$~GeV$^{-1}$
was obtained for $m \lesssim 0.02$ eV~\cite{cast2007}. This is comparable
to the astrophysical limit of $g_{{a\gamma}}< 6 \times 10^{-11}$~GeV$^{-1}$
from HB stars in globular clusters \cite{Raffeltbook}. Solar axions
experiments do not rule out the parameter region where $m \gtrsim 0.02$ eV,
but they provide one of the best limits at the moment. Another
experimental limit comes from axion dark ma\-tter searches employing
microwave cavities. The ex\-pe\-ri\-ment works under the assumption that
if axions comprise most of the dark matter in galaxies their mass
should have a lower bound $m > 10^{-6}$ eV, otherwise they could,
probably, overclose the universe \cite{sikivie-cosmo}.
Although these experiments are restricted to a narrow  mass interval
they test models with $g_{{a\gamma}}> 10^{-15}$~GeV$^{-1}$
for $1.9 \times 10^{-6} < m < 3.3 \times 10^{-6}$ eV \cite{admx_2004}.

The astrophysical bounds, although robust,
are model-dependent and may be relaxed by many orders of magnitude \cite{alt1,moha}.
In particular, it is widely assumed that the pseudoscalar mass and the pseudoscalar-photon
coupling constant are unaffected by environmental effects such as the temperature
or the energy density of the stellar plasma.
Thus, purely laboratory-based searches for pseudoscalar particles are important because
they provide a fully model-independent approach. They do not rely
on the physical processes and conditions in astrophysical
environments neither on how axion-like particles could be produced
under these conditions.

In general, laser based experiments can be
divided into two categories: the so called \textit{photon
regeneration} or \textit{light shining through a wall}  experiments
\cite{vanB87} and experiments that probe the  magneto-optical
properties of the vacuum \cite{asztalos-axionrev06}. We shall deal here with photon regeneration
experiments. The basic idea is the following: a polarized laser beam propagates
inside a transverse magnetic field in order to induce the
transmutation of some photons into axion-like particles. At some
point of its path light is blocked by a wall, such that only the
axion-like particles created before will be able to pass through it.
A second magnetic field behind the wall is used to reconvert the
axions to photons which are finally detected. The pioneering experiment
based on this technique was performed at the beginning of the
1990s \cite{ruoso-cameron_92}. Present limits on the axion
parameters are still less stringent that the limits obtained by solar
searches (see e.g. \cite{robilliard-2007,osqar}) but several improvements
have been proposed that would allow to probe values of $g_{{a\gamma}}$ not previously
excluded by stellar evolution limits or solar axion searches
\cite{sikivie-PRL}.

\begin{figure}
\includegraphics[angle=-90,width=9cm]{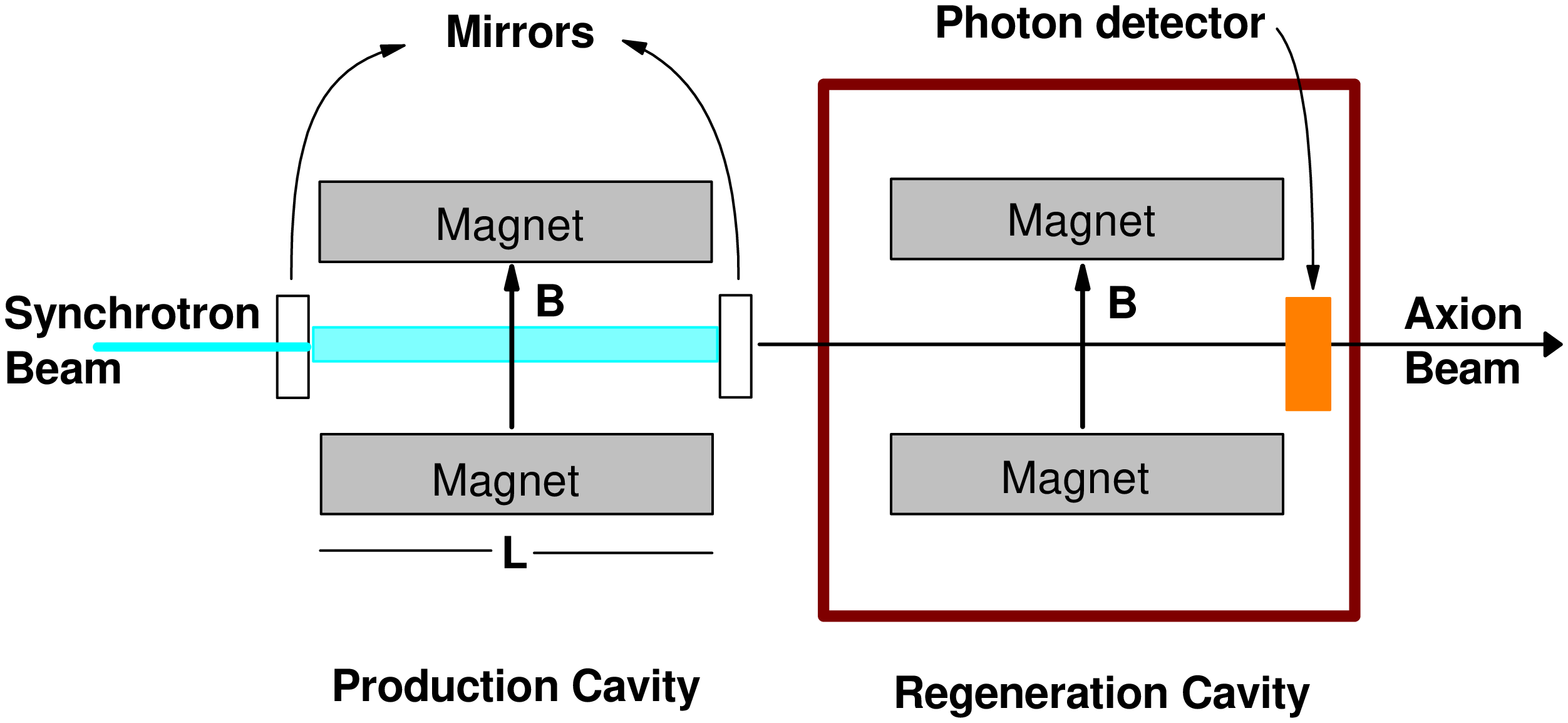}
\caption{Simple photon regeneration experiment using synchrotron
light. The photon beam may be enhanced in the first cavity by producing
several reflections inside the first magnet like in the laser
photon-regeneration experiment performed by Ruosso et al.
\cite{ruoso-cameron_92}. }
\label{design}
\end{figure}

In this work we propose a photon regeneration experiment similar
to the one described above but employing synchrotron light
instead of a laser beam (see Fig. \ref{design}). The synchrotron
photon beam travels inside a pseudoscalar ``production" cavity
immersed in an transverse magnetic field  in order to induce the
hypothetic conversion of a fraction of the photons into pseudoscalars.
These particles may be converted back into detectable photons in the
``regeneration" cavity immersed in a second magnetic field B of the
same intensity.
In the following we shall assume a device utilizing in each cavity
a number $N_m$ of superconducting magnets like those employed
in the Large Hadron Collider (LHC), i.e. having a longitude of $L \sim 15$ m and a
magnetic field strength $B \sim 10$ T. Additionally, the number
of conversion events may be enhanced by introducing a pair of mirrors
in the ``production'' cavity causing the photon beam to traverse
the magnet several times before exiting.
This is an experimental challenge, since it is necessary
to construct high reflectivity mirrors for frequencies
ranging from the visible part of the spectrum to hard
X-rays (see Discussion). Nonetheless, we shall explore
this possibility as well. In the following we analyze the formula
for the probability of the
process $\gamma \rightarrow a  \rightarrow  \gamma $ for monochromatic
photons and then we calculate the number of regeneration events that
may be expected in the experiment using a beam of synchrotron light.

Before addressing the conversion probability for the proposed experiment,
we point out that there are other motivations for researching on very light and very
weakly interacting particles along the lines discussed here.
For example, developments in string theory show that an
effective Lagrangian like Eq. (\ref{Leff}) is a natural outcome of the theory,
although one needs extra assumptions to make such string axion models
to have an $f_a$ scale compatible with the usual cosmological arguments
\cite{svrcek-witten} (see also \cite{ringwald2008}).
There are also string based models with millicharged fermions, arising from particular
compactifications of extra dimensions leading to hidden sector U(1) gauge factors, which
mix with the usual U(1)$_Y$ hypercharge factor through kinetic terms \cite{abel-et.al-2006}.
These models could be constrained in current and near future laboratory experiments
similar to the one proposed here \cite{abel-et.al-2006}. This emphasizes the connection
of low energy phenomena involving very light particles with other
theoretical and experimental activities in particle physics \cite{ringwald2008}.

\textit{Probability of conversion $\gamma\rightarrow a\rightarrow\gamma $: }
In the case of very relativistic pseudoscalars  ($m \ll \omega$), the
short-wavelength approximation can be applied and the equations of
motion reduce to a first order propagation equation. Considering a
monochromatic light beam traveling along the $z$-direction in the
presence of a transverse homogeneous magnetic field $B$, it can be
shown that the probability  for a photon to convert into a
pseudoscalar after traveling a distance $s$ is given by
\cite{Raffeltbook}
\begin{equation}
 P_{\gamma a}(\omega)  = \mathcal{A}^2(\omega) \sin^2
 \left[ \frac{\beta}{ \mathcal{A}(\omega)}  \right] \, ,
\label{pagamma1}
\end{equation}
where the dimensionless quantities $\beta$ and $\mathcal{A}$ are defined as:
\begin{eqnarray}
\beta \equiv \frac{g_{a \gamma} B s}{2},  \;\;\;\;\;\;
\mathcal{A}^2(\omega)  \equiv  \frac{1}{1+(\omega_{*}/\omega)^2},
\nonumber
\end{eqnarray}
and we introduced a critical energy:
\begin{eqnarray}
\omega_{*} \equiv\frac{m^2}{2 g_{a \gamma} \,B} \, .\nonumber
\end{eqnarray}
The plasma and vacuum polarization effects are being disregarded
since they do not play any dominant role in the experiment we are
considering. We are also using a different parametrization for
the probability which is usually presented in its original general form,
in terms of the mixing angle $\theta$ and oscillation length $l_{osc}$
as in \cite{Raffeltbook}. By a direct identification
it is seen that $\sin 2 \theta=\mathcal{A}(\omega)$ and
$l_{osc}= 2\pi\mathcal{A}(\omega)/g_{a \gamma} \,B$.

As stated above, we shall assume a device utilizing a number $N_m$ of
LHC-like magnets, thus  $s = N_m L \sim  N_m \times 15$ m and $B \sim 10$ T.
For these typical values we
can parameterize the critical energy $\omega_{*}$ and the factor
$\beta$ as:
\begin{equation}\label{cond1}
\omega_{*}   =  2.56 \, \mathrm{keV} \times \frac{ \: m_{\mu {\rm
eV}}^2}{g_{10}\,B_{\rm 10T}}\, ,
\end{equation}
\begin{equation}
\beta = 0.74 \times 10^{-8}  g_{10} B_{10T} N_m L_{15m},
\end{equation}
where $m_{\mu \mathrm{eV}} = m / (10^{-6} \mathrm{eV})$,
$g_{10}= g_{a \gamma}/ (10^{-10} \mathrm{GeV^{-1}})$,
$B_{10T}= B / (10 \mathrm{T})$ and  $L_{15m} = L / (15 \mathrm{m})$.

Some limits of the probability given by Eq. (\ref{pagamma1})
are of interest. Depending on the relative values of $g_{a \gamma}$
and $m$, the photon frequencies employed in the experiment may be
larger or smaller than the critical frequency $\omega_{*}$.
For $\omega \gg \omega_{*}$ we have $\mathcal{A} \sim 1$. Since the
factor $\beta$ given above is $\ll 1$ for the typical values employed
here, we can approximate the sine by its argument resulting:
\begin{equation}
 P_{\gamma a} \approx \beta^2
\label{approx1}
\end{equation}
which is independent of $\omega$. For $\omega \ll \omega_{*}$
we have $\mathcal{A} \approx 2 g_{a \gamma} B \omega / m^2  \ll 1$.
If $N_m L \ll 4 \omega / m^2$ (i.e. $\beta  \ll \mathcal{A}$) we
arrive again to Eq. (\ref{approx1}). However, if $N_m L \gg 4 \omega / m^2$
(i.e. $\beta \gg \mathcal{A}$), we can approximate the sine squared
by $1/2$, resulting:
\begin{equation}
 P_{\gamma a}  \approx \frac{\omega^2}{2\omega_{*}^2} =
 \frac{2 g_{a \gamma}^2 B^2\omega^2 }{ m^4}\ll \beta^2.
\label{approx2}
\end{equation}
In this limit, $P_{\gamma a}$ depends on $\omega$ and on the
mass of the pseudoscalar particle.

\begin{figure}
\includegraphics[angle=-90,width=9cm]{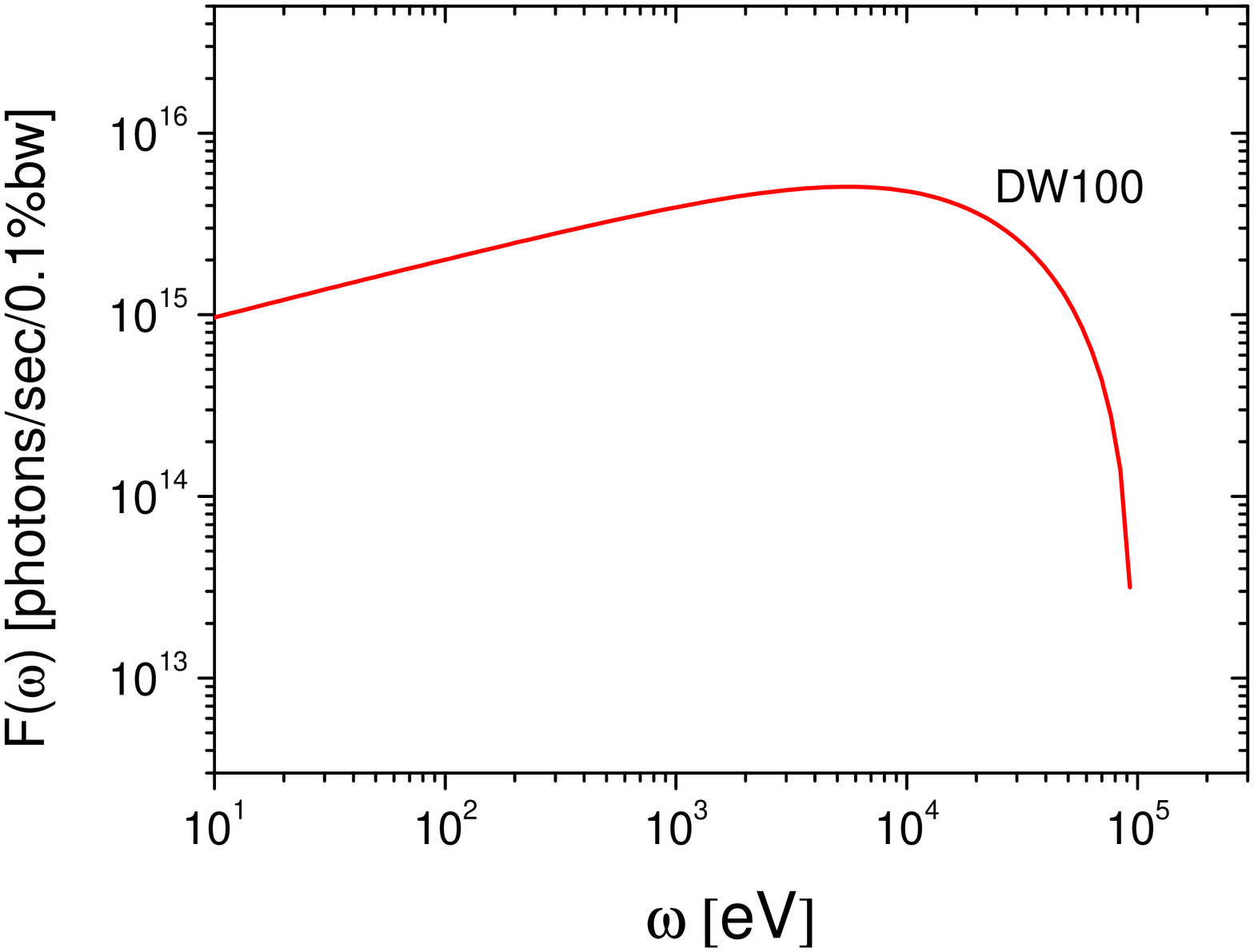}
\caption{Flux versus photon energy for the typical synchrotron device
considered in this work. The flux is defined through
$F(w) \equiv  {\omega \; n(\omega)}/{1000}$ and is given in units
of photons per second per $0.1 \%$ band width \cite{NLSL}.  }
\label{fig-flux}
\end{figure}

\textit{Conversion of a synchrotron beam:}
The probability  $\gamma\rightarrow a\rightarrow\gamma $  given above must be
integrated over all frequencies in the synchrotron beam in order
to calculate the number of  events in the regeneration cavity.
On the other hand, the pseudoscalar production rate can be increased by
reflecting light several times inside the first magnet,
thus increasing the power of the synchrotron beam.
This can be done by using an ``optical delay line'', i.e.
an incoherent cavity encompassing the production magnet,
as reported e.g. in Ref. \cite{ruoso-cameron_92}.
If the reflectivity of the mirrors is $R = 1 - \tilde{\eta}$,
the power of the right-moving photon-beam is increased
approximately by a factor $1/(2\tilde{\eta} - \tilde{\eta}^2 )\equiv 1/\eta $, and so is the spectral
density $n(\omega) \equiv d^2N / d\omega dt$ (number of
photons in the beam per unit frequency and unit time).
Notice that in general $\eta$ would be a strongly dependent
function of $\omega$. However,  we shall consider $\eta$
as a constant in order to determine whether the optical delay line
can produce a significant enhancement of the regeneration rate (see below).
In the production cavity, the number of photons that
would be converted into pseudoscalars per unit time is
\begin{eqnarray}
\frac{dN_{\gamma a}}{dt} = \int_{\omega_1}^{\omega_2}
P_{\gamma a}(\omega) \frac{n(\omega)}{\eta} d\omega,\nonumber
\end{eqnarray}
with the beam having a significant flux between the frequencies
$\omega_1$ and $\omega_2$. The spectral density
in a synchrotron machine can be approximated by

\begin{equation}
n(\omega) = \mathcal{F}_0 \int_{\omega/\omega_c}^{\infty}
K_{5/3}(x) dx  \; ,
\end{equation}
where $\mathcal{F}_0$ is a constant, $\omega_c$ is the critical
synchrotron frequency below which the spectrum of the radiation will
contain appreciable frequency components, and  $K_{5/3}(x)$ is the
modified Bessel function \cite{Jackson}. For example, in  a synchrotron
facility like the Damping Wiggler DW100 at NSLS-II the spectral flux
density is such that $\mathcal{F}_0 \sim 10^{15} \mathrm{ph/s/eV}$
with $\omega_c \sim 2 \times 10^4 \mathrm{eV}$ (see Fig. \ref{fig-flux}).

The pseudoscalar beam coming from the production cavity
would cross the wall and would have the probability
$P_{a \gamma}$ to be reconverted into photons in the regeneration
cavity. Taking, for simplicity,  the same geometry and magnetic
field configuration as in the first cavity we have
$P_{a \gamma} = P_{\gamma a}$.
On the other hand, the detection of the regenerated photons is affected by the efficiency $\epsilon$ of the photon detectors.
X-ray detectors like those widely used in X-ray astronomy have efficiencies close to 1 in the range of $10^2-10^5$ eV (see e.g. \cite{xrays}). Efficiency in the ultraviolet energy range is smaller, from several
percent to tens of percent. However, as seen from Fig. 2 this part of the spectrum contributes with
a small fraction of the total synchrotron photon-flux ($\sim 10 \%$) and thus it
is not determinant for the results.
For simplicity, we conservatively assume $\epsilon = 0.5$ for all frequencies.
The number of regeneration events $\gamma  \rightarrow a  \rightarrow \gamma$  in an operation
time $\Delta t$ is then:
\begin{eqnarray}
N_{\gamma a \gamma}  =
\Delta t  \times \int_{\omega_1}^{\omega_2}
P^2_{\gamma a}(\omega) \frac{\epsilon \; n(\omega)}{ \eta} d\omega.
\label{Ngeral}
\end{eqnarray}

According to previous experiments, the probability for the process
$\gamma \rightarrow a  \rightarrow  \gamma $ should be quite small,
if it is not zero. We shall consider a null experiment (no detection
after an observation time $\Delta t$) in order to determine the exclusion
limits on $g_{a \gamma}$ and $m$ to the 95\% CL. Assuming a Poisson
distribution, this corresponds to taking $N_{\gamma a \gamma}  = 3$
in Eq. (\ref{Ngeral}).

\begin{figure}
\includegraphics[angle=-90,width=10cm]{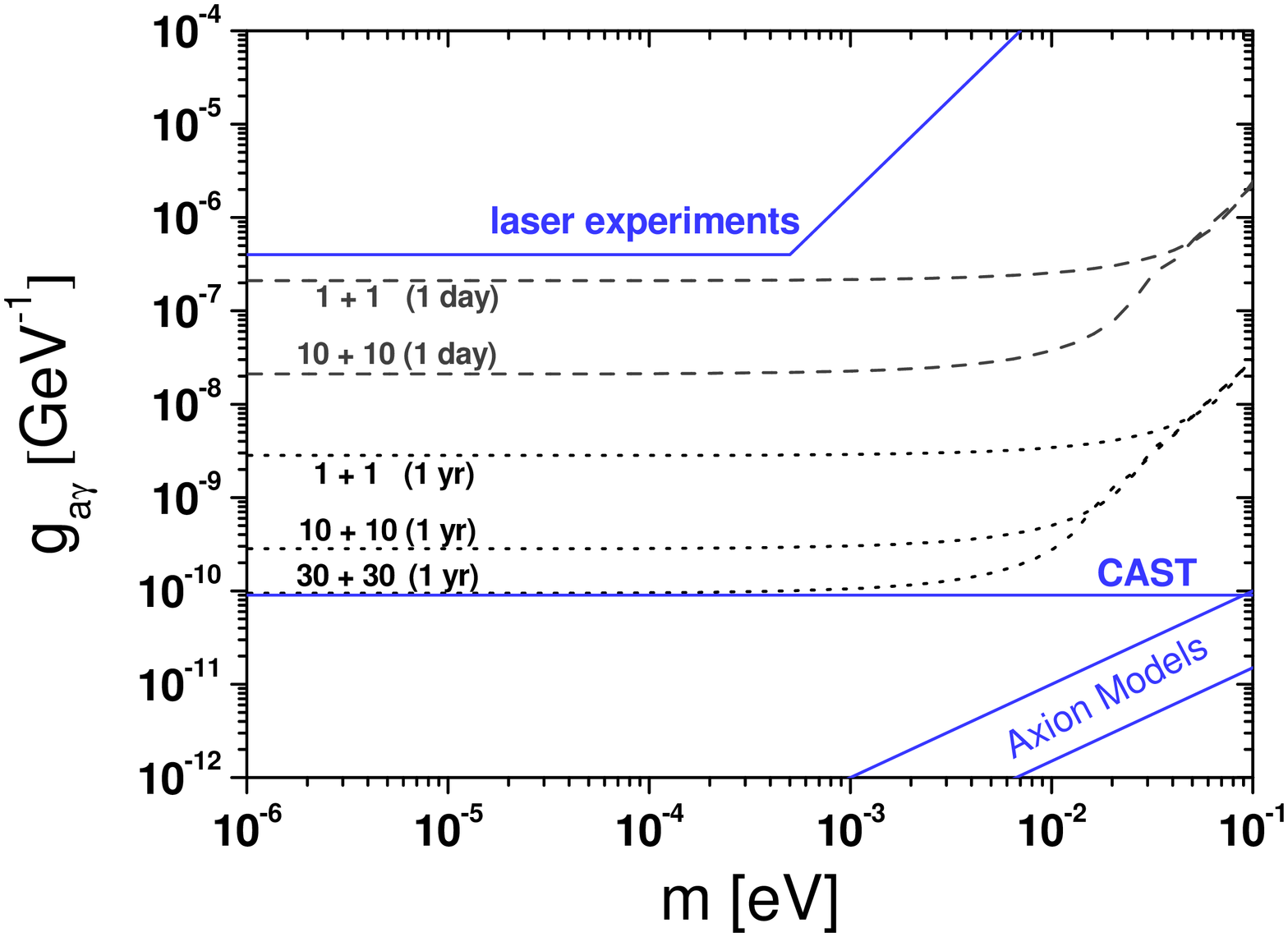}
\caption{We show the $95 \%$ confidence level exclusion limits on
$g_{a \gamma}$ and $m$ for configurations using $N_m + N_m$ magnets
after one day and one year of cumulative running, adopting $\eta = 1$ and $\epsilon = 0.5$.
We also show the present exclusion line from CAST  \cite{cast2007} and from laser experiments \cite{ruoso-cameron_92,robilliard-2007}, and the region
corresponding to axion models according to Eq. (3).  Notice the significant improvement
in the mass reach of the synchrotron radiation experiment in comparison with laser experiments.
This is due to the use of a broadband incident photon-beam with higher frequencies than that
of infrared or optical lasers.}
\label{fig-results-1}
\end{figure}

Let us analyze some limits of Eq. (\ref{Ngeral}). When $\omega \gg \omega_{*}$,  $P_{\gamma a}(\omega)$ is given
by Eq. (\ref{approx1}) and therefore
$N_{\gamma a \gamma}  = \Delta t \times \beta^4 N_{ph} \epsilon / \eta$, where
$N_{ph}=  \int_{\omega_1}^{\omega_2} n(\omega) d\omega$ is the total
number of photons per second in the synchrotron beam. Thus, we may write
\begin{eqnarray}
g_{10} \approx 18
\bigg[ \frac{\eta N_{\gamma a \gamma} }{ \epsilon \Delta t_{yr} N_{ph20}} \bigg]^{1/4}
\frac{1}{B_{10T} \, N_m \, L_{15m}} ,
\label{limit-large}
\end{eqnarray}
where $\Delta t_{yr} \equiv \Delta t / (1 \; \mathrm{year})$  is the
cumulative running time in units of one year,
$N_{ph20} \equiv N_{ph} / (10^{20} \mathrm{photons \times s^{-1}})$
is parameterized for the typical values corresponding to the DW100
shown in Fig. \ref{fig-flux}, and $N_m$ is the number of identical
LHC-like magnets employed in each cavity of Fig. 1. Notice that Eq.
(\ref{limit-large}) does not depend on $m$.

In the limit $\omega \ll \omega_{*}$ with
$N_m L \gg 4 \omega / m^2  $, the probability
$P_{\gamma a}(\omega)$  is given by Eq. (\ref{approx2}) and
$N_{\gamma a \gamma}$ reads:
\begin{eqnarray}
N_{\gamma a \gamma}  =  \Delta t \times
\frac{4 \,  \epsilon \, g_{a \gamma}^4 B^4}{\eta \, m^8} \int_{\omega_1}^{\omega_2}
\omega^4 n(\omega)   d\omega
\end{eqnarray}
Evaluating numerically the above integral in the case of the DW100 we have:
\begin{eqnarray}
g_{10} \approx \frac{  2 \times 10^{-8}}{B_{10T}}
\bigg[ \frac{\eta \, N_{\gamma a \gamma} }{\epsilon \Delta t_{yr}} \bigg]^{1/4}
m^2_{\mu {\rm eV}},
\label{limit-small}
\end{eqnarray}
which is independent of the number and longitude of the magnets.
In this case the experiment's reach in $g_{a \gamma}$ degrades considerably
as the mass increases. Notice that in both Eqs. (\ref{limit-large})
and (\ref{limit-small}) the limit on the coupling
constant $g_{a \gamma}$ scales with $\eta^{1/4}$ and with $\Delta t^{-1/4}$.
Thus,  the most efficient way to improve the limits is to increase the
magnetic field and/or the number of magnets $N_m$. Also, since the efficiency of the photon detector enters trough $\epsilon^{1/4}$, there are no strong differences between the results
in the ideal case with $\epsilon = 1$ and a realistic case with  $\epsilon \approx 0.5$.
In Fig. \ref{fig-results-1} we show the $95 \%$ C.L.
limits on $g_{a \gamma}$ and $m$ calculated by solving numerically
Eq. (\ref{Ngeral}) for different cumulative running times and different
number of magnets in each cavity. For small enough values of $m$
most of the frequencies in the synchrotron beam are such that
$\omega \gg \omega_{*} \equiv {m^2}/({2 g_{a \gamma} \,B})$.
In this case the probability $P_{a \gamma}$ is independent of $m$ and thus,
the horizontal lines of Fig. \ref{fig-results-1} are obtained (in coincidence with Eq. (\ref{limit-large})).
For large  enough values of $m$  most of the
frequencies verify $\omega \ll \omega_{*}$. In this case the
asymptotic straight lines of Fig. \ref{fig-results-1} are given by
Eq. (\ref{limit-small}).

\textit{Discussion and conclusions:}
In this paper we have shown that a photon regeneration
experiment using synchrotron light can be used to
set limits on $g_{a \gamma}$ and  $m$ thanks to the high
photon flux of the device.
In fact, with a configuration
of 1 + 1 LHC magnets and one day of cumulative
operation it is possible to improve the present limits on $g_{a \gamma}$ imposed by laser experiments
(for these experiments present limits are roughly
$g_{a \gamma} \lesssim 4 \times 10^{-7}$~GeV$^{-1}$
\cite{ruoso-cameron_92,robilliard-2007} in the range of $m \lesssim 10^{-4}$ eV ).
Moreover, there is a significant improvement in the sensitivity of the experiment
for $m \gtrsim 10^{-4}$ eV.
For a laser-experiment and a synchrotron-experiment having the same sensitivity in the
range of $m \lesssim 10^{-4}$ eV, the synchrotron one is three orders of magnitude better
for masses around  $10^{-2}$ eV (see Fig. \ref{fig-results-1}). This is due to the use of a broadband incident photon-beam with higher frequencies than that of infrared or optical lasers.
The limits of the  1 + 1 (one day) configuration are improved by more than one order
of magnitude in the case of a 10 + 10 configuration working for
1 day, and by more than two orders of magnitude in the case of
a 1 + 1 configuration working for one year. In the case
of a 10 + 10 configuration working for one year the obtained limits
are three orders of magnitude better and very close to the limits
obtained by CAST and from HB stars.
For completeness, we notice that the configuration that is necessary in order to improve in a
quite model-independent way the
limit $g_{a \gamma} < 6 - 9 \times 10^{-11}$~GeV$^{-1} $  imposed by
solar searches and HB stars in globular clusters consists of  30 + 30 LHC magnets
working for one year (see Fig. \ref{fig-results-1} and Eq. (\ref{limit-large})).
The latter is quite ambitious and expensive but not impracticable (notice that
approximately 1600 magnets are employed at the LHC).

Since the proposed experiment is expected to work at an existing synchrotron source,
most of the cost is related to the superconducting magnets (for a recent detailed
study of costs of high-field superconducting strands for particle accelerator magnets
see \cite{cost}). On the other hand, costs may be significantly reduced by re-using
superconducting prototypes of large experiments. An interesting example is the OSQAR
experiment \cite{osqar} which re-uses superconducting dipoles developed at CERN for the Large Hadron Collider.

Notice that pseudoscalars are (potentially) produced if the electric field of the photons and the magnetic field of the magnets are oriented parallel to each other (${\cal L}_{{a\gamma}}=  g_{{a\gamma}}\, a\, {\bf E \,.\, B}$). Light leaving damping wigglers (like DW100 at NSLS-II) is usually polarized in the horizontal plane while the proposed LHC magnets have a perpendicular magnetic field. This means that the magnets should be rotated by 90 degrees with respect to their standard orientation.
We also notice that it is not necessary to tie up a beam line exclusively for this experiment.
As emphasized by Rabadan et al. \cite{raba2005}, if the initial conversion magnet
is placed before a target that is the subject of other experiments, it is
possible to perform both experiments simultaneously since the pseudoscalar
beam will propagate unimpeded through the target.

Concerning the background signal the situation should not be very
different than for ``light shining through a wall'' experiments using
optical \cite{osqar} or X-ray \cite{raba2005} laser beams. The background signal
should come mainly from the CCD readout noise and
from cosmic rays, both of which can be controlled by standard procedures
(see e.g. \cite{osqar,raba2005}).

Also, as for laser experiments, the number of conversion
events could be enhanced by introducing a pair of mirrors in the
``production'' cavity causing the photon beam to traverse the
magnet several times before exiting. However, notice that the limit on
$g_{a \gamma}$ scales with $\eta^{1/4}$ (see Eqs. (\ref{limit-large})
and (\ref{limit-small})). Thus, ultra-high-reflectivity mirrors would be
necessary in order to increase significantly the limits on $g_{a \gamma}$.
For synchrotron light this is a huge experimental challenge,
since it is necessary to construct mirrors with good reflectivity
for frequencies ranging from 10 eV to 100 keV. Reflectivities
are good in the visible spectral region but are still very poor
for larger frequencies (e.g. for X-rays it is $R \lesssim 80 \% $).
Moreover, X-rays can be reflected off smooth
metallic surfaces only at very shallow angles. In fact, the
difficulties in constructing normal incidence x-ray mirrors of
high reflectivity is one of the main technical challenges in the
realization of x-ray resonators. Some experiments have shown
that a Fabry-Perot resonator for hard X-rays and near gamma-rays
could be realized in principle \cite{FP}. This could be interesting
in view of the recent paper of Sikivie et al. showing that
photon-regeneration experiments may be resonantly enhanced by
employing matched Fabry-Perot cavities encompassing both cavities
of Fig. 1 \cite{sikivie-PRL}. While this could be very useful for
the recently proposed regeneration experiment using x-rays from a
free-electron laser \cite{raba2005}, it is not clear whether it
could ever be implemented for the broad-band beam of a synchrotron source.


We emphasize that, although  the more realistic configurations of the proposed experiment
explore a region of the parameter space with larger values of
$g_{a \gamma}$ than the explored by solar experiments and HB stars,
it is important to test the parameter space in the laboratory
since there could be unknown mechanisms suppressing the pseudoscalar
emission in astrophysical sources \cite{alt1,moha}.
For example, the CAST experiment \cite{cast2007} looks for
pseudoscalar particles from the Sun produced by the interaction of photons
with plasmons in the solar plasma:  $\gamma +  \mathrm{plasmon} \rightarrow a$. However, the production
of axions through this reaction may be prevented by some sort of mechanism.
In fact, there are models where the effective Lagrangian has the interaction
$\phi a F \tilde{F}/M^2$ in place of $a F \tilde{F}$, where $\phi$ is a light
scalar field which can affect the pseudoscalar production dynamics at high temperatures \cite{moha}.
Assuming some properties for $\phi$ it can be shown that pseudoscalar production
is suppressed for temperatures $T \gg \mathrm{keV}$ as in the Sun,
but allowed in the laboratory \cite{moha}. The same argument
holds for pseudoscalar particles produced in HB stars.
This exemplifies the need for purely laboratory experiments
since we cannot guarantee that the interaction
in high temperature environments is the same as at very low temperatures.
Additionally, experiments like the one proposed here
may help to broaden the research field from searches for just axion-like particles to other very light and very weakly interacting particles arising in string-theory motivated extensions of the Standard Model.

This work was supported by State of S\~{a}o Paulo Research
Foundation (FAPESP). A. G. D. and  G. L. also thank CNPq (Brazil) for financial support.
We acknowledge the referee for helpful comments.

\end{document}